\documentclass{aa}  
\usepackage{color}
\usepackage{graphicx}

\newcommand{\HI}{{\ion{H}{1}}}

\newcommand{\kms}{$\,$km$\,$s$^{-1}$}

\newcommand{\msun}{{$M_\odot$}}

\newcommand{\co}{{CO(2-1)}}
\def\HI{\ion{H}{i}}

\def\emph#1{{\sl #1}}
\newcommand{\ltsima} {$\; \buildrel < \over \sim \;$}
\newcommand{\gtsima} {$\; \buildrel > \over \sim \;$}
\newcommand{\lta} {\lower.5ex\hbox{\ltsima}}
\newcommand{\gta} {\lower.5ex\hbox{\gtsima}}

\usepackage{txfonts}
\begin{document} 

\title{The missing link: Tracing molecular gas in \\the outer filament of Centaurus A}
\authorrunning{Morganti et al.}
\author{Raffaella Morganti\inst{1,2}, Tom Oosterloo\inst{1,2}, J. B. Raymond  Oonk\inst{1,3},\\ Francesco Santoro\inst{1,2}  \and  Clive  Tadhunter \inst{4}}

\institute{ASTRON, the Netherlands Institute for Radio Astronomy, Postbus 2, 7990 AA, Dwingeloo, The 
Netherlands. \\
 \email{morganti@astron.nl}
\and
Kapteyn Astronomical Institute, University of Groningen, P.O. Box 800,
9700 AV Groningen, The Netherlands
\and
Leiden Observatory, Leiden University, P.O. Box 9513, 2300 RA Leiden
\and
Department of Physics and Astronomy, University of Sheffield, Sheffield, S7 3RH, UK
}

 
  \abstract
   {We report the detection, using observations of the \co\ line performed with the Atacama Pathfinder EXperiment (APEX), of molecular gas in the region of the outer filament  of Centaurus A, a complex region known to show various signatures of an interaction between the radio jet, an \HI\ cloud,  and ionised gas filaments. We detect \co\ at all  observed locations, which were selected to represent regions with very different physical conditions.    
The H$_2$ masses of the detections range between $0.2 \times 10^6$ and $1.1 \times 10^6$ \msun, for conservative choices of the CO to H$_2$ conversion factor.
Surprisingly, the stronger detections  are not coincident with  the \HI\ cloud, but instead  are in the region of the ionised filaments. We also find variations in the widths of the \co\ lines throughout the region, with broader lines in the region of the ionised gas, i.e.\  where the jet--cloud interaction is strongest, and with narrow profiles in the \HI\ cloud. 
This may indicate that the molecular gas in the region of the ionised gas has the momentum of the jet--cloud interaction encoded in it, in the same way as the ionised gas does. These molecular clouds may therefore be the result of very efficient cooling of the down-stream  gas photo- or shock-ionised by the interaction. On the other hand, the molecular clouds with narrower profiles, which are closer to or inside the \HI\ cloud, could be pre-existing cold H$_2$ cores which manage to survive the effects of the passing jet.

}

   \maketitle  
%

\section{Introduction}

Radio-loud active galactic nuclei (AGN) are known to inject energy into the surrounding interstellar medium (ISM)  via plasma jets. The impact this has on the host galaxy is relevant on both  large and small scales (see e.g.\ \citealt{McNamara12,Morganti13}). The induced compression -- and subsequent cooling -- of gas disturbed by the transit of a radio jet can  induce the formation  of new stars (e.g.\ \citealt{Breugel93,Dey97,Oosterloo05,Croft06}), but AGN-driven gas outflows can also occur which  remove, or at least redistribute, the cold gas and/or create highly turbulent conditions which inhibit star formation (e.g. \citealt{Alatalo11,Combes13,Burillo14,Morganti13,Guillard15,Morganti15}).

One of the important results of recent work is that molecular gas often is the most massive component in jet-cloud interactions  (e.g.\ \citealt{Feruglio10,Alatalo11,Cicone14,Morganti15}). 
Thus, tracing this phase of the gas  provides key information on how these processes work and how they depend on the properties of the gas and  of the radio jet. 

When it comes to studying the interaction between a radio jet and the ISM of the host galaxy, the nearest radio-loud AGN, Centaurus A (Cen A, $D$ = 3.8 Mpc, \citealt{Harris10}), should be considered a prime target. In this galaxy, about 15 kpc NE from its centre, the so-called \emph{outer filament} region  presents an extremely intriguing situation. Past studies of this region have revealed many signatures of an ongoing interaction between the radio jet and gas clouds (see \citealt{Morganti10} for a review) and over the years a wealth of data has been collected on this region, covering almost all wavebands. However, although it is  such a well-studied region,  very little information is available about the molecular gas in and around the outer filament, and this has motivated the \co\  observations with the Atacama Pathfinder EXperiment (APEX)  presented in this paper. 

\section{Special region: The    outer filament}

The outer filament was  discovered as a filament of highly ionised gas outside the optical body of Cen A,  well aligned with the radio jet of Cen A \citep{Blanco75,Graham81,Morganti91,Morganti99,Morganti10}. Later work identified regions of ongoing star formation   in  the filament \citep{Mould00,Rejkuba01}, while 
bright UV emission was detected with GALEX \citep{Neff15}. 
About 2 kpc west of the ionised gas, a large \HI\ cloud was found \citep{Schiminovich94}, likely the remnant of a major accretion by Cen A. In this \HI\ cloud, at locations closest to the radio jet, \HI\ with velocities very different from those of the regular rotation of the \HI\ cloud around Cen A was detected \citep{Oosterloo05}, which was taken as evidence that   the radio jet  is affecting the \HI\ cloud at those locations. This is further confirmed by integral field spectroscopy with VIMOS and MUSE \citep{Santoro15a,Santoro15b}, showing the presence of even more disturbed kinematics in the ionised gas. The MUSE data reveal three morphologically and kinematically distinct  components in the ionised gas, which are thought to correspond to different stages in the jet-cloud interaction. The data also show that, overall, the ionisation of the gas is due to ionising photons from the AGN, but that locally star formation also plays a role \citep{Santoro16}. Interestingly, the rate at which the available gas reservoir is turned into stars is low, possibly connected to increased turbulence powered by injection of kinetic energy by the jet \citep{Salome16}. 
The Balmer decrement (H$\alpha$/H$\beta$) across the MUSE fields confirms that the ionised gas is dusty. 
The presence of  very cold dust ($T$$\sim$13 K)   in and around the outer filament  is also seen in Herschel data \citep{Auld12}. 

The above  shows that the region around the outer filament is very rich in phenomena and that it shows the full complexity of the interaction of a jet  with a large gas cloud. This interaction is clearly stirring up a gas cloud and is  destroying it (partially), but at the same time star formation is happening with low efficiency.

To complete this picture, information on a key component is missing: the molecular gas. The only observations available around this region are those done with SEST by \citet{Charmandaris00} of locations in the \HI\ cloud and ALMA observations of a single location just north of the \HI\ cloud \citep{Salome16}. They have detected molecular gas associated with the northern region of this cloud and estimate that the molecular gas is the most massive component at the observed locations. These observations, however, only  sample regions relatively distant  
($\sim$3 kpc) from the outer filament and from the region with anomalous \HI\ velocities. Here we report observations which have a more complete coverage of the outer filament.

\begin{table}[b] 
\caption{Locations of the APEX pointing positions and the noise of the spectra.}
\centering 
\begin{tabular}{cccc} 
\hline\hline 
   Position & R.A.\ (J2000) & Dec (J2000) & Noise  \\
    & (h m s) & (d m s)  & (mK)\\
\hline
 F2       & 13 26 21.9   &  $-$42 48 58.5 & 3.2\\
 F3             & 13 26 13.1   &  $-$42 50 03.1 & 1.8\\
 F4             & 13 26 18.4   &  $-$42 49 18.1 & 1.2        \\
 F5             & 13 26 16.4   &  $-$42 47 59.5 & 3.3 \\
 F6                     & 13 26 28.8   &  $-$42 49 52.5 & 2.5 \\
\hline\hline 
\end{tabular}
\label{tab:obs}
\end{table}

   \begin{figure}
   \centering
    \includegraphics[width=8cm]{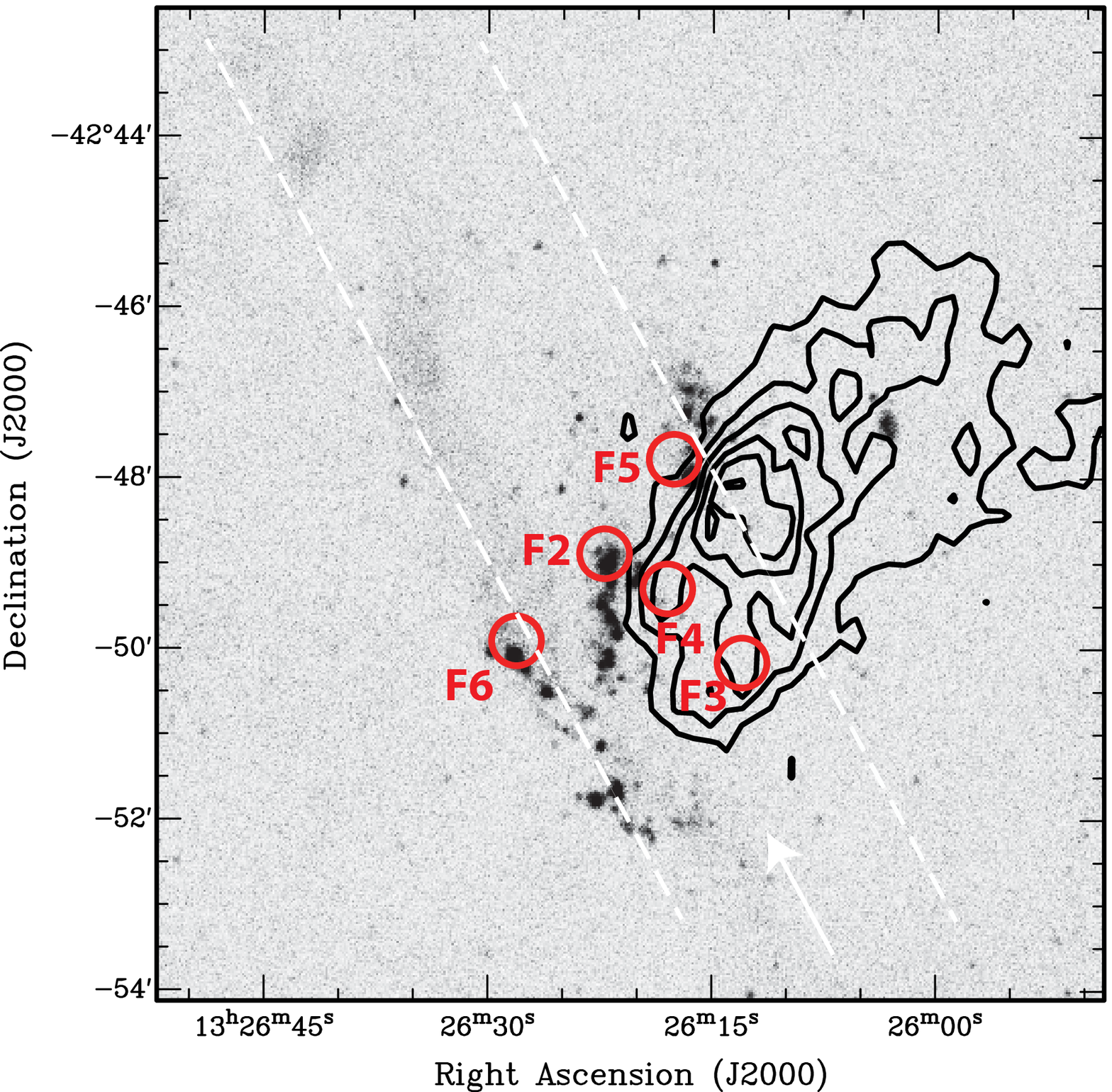}
   \caption{Locations of the APEX pointings (see Table \ref{tab:obs}) on the GALEX FUV image of the outer filament \citep{Neff15}. The black contours give the distribution of the \HI\ cloud with contour levels 1, 4, 7, 10, 13, and 16 $\times$ $10^{20}$ cm$^{-2}$  \citep{Oosterloo05}. The dashed lines  approximately  delineate the path of the radio jet \citep{Morganti99}; the arrow shows the flow direction of the jet.}
              \label{fig:pointing}
    \end{figure}

   \begin{figure}
   \centering
 \includegraphics[width=8.5cm,angle=0]{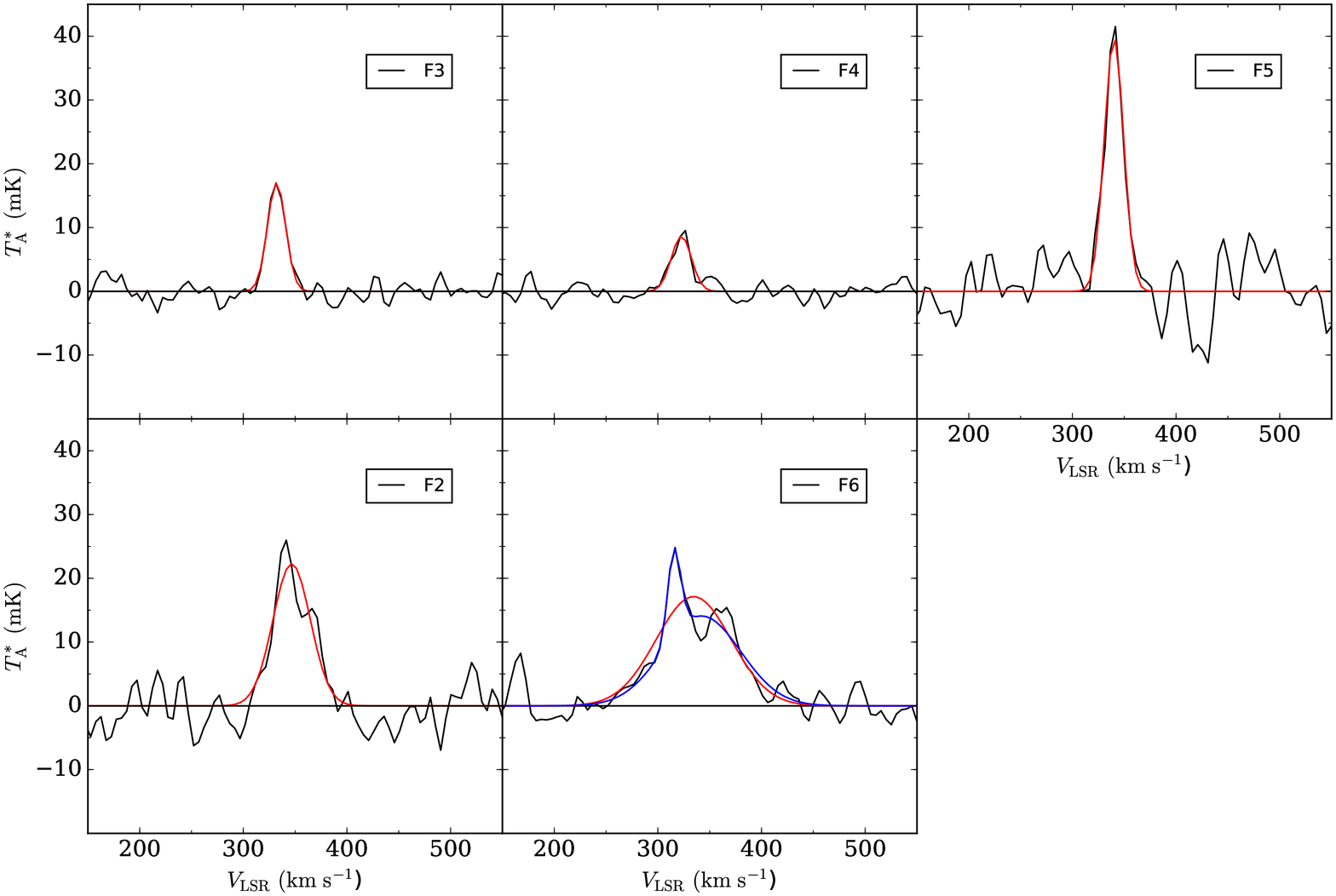}
   \caption{\co\ spectra for the observed locations. Red lines show the single-component Gaussian fits, while the blue line for F6 shows the double-Gaussian fit.}
              \label{fig:profiles}%
    \end{figure}
%

\begin{table*}
\centering
\begin{tabular}{c r@{ $\pm$ }l r@{ $\pm$ }l r@{ $\pm$ }l  r@{ $\pm$ }r  r@{ $\times$ }l c c c}
\hline
\hline
Position
& \multicolumn{2}{c}{Peak}
& \multicolumn{2}{c}{Velocity}
& \multicolumn{2}{c}{Dispersion}
& \multicolumn{2}{c}{Integrated flux}
& \multicolumn{2}{c}{H$_2$ mass}
& H$_2$/\HI\ 
 \\
& \multicolumn{2}{c}{(mK)}
& \multicolumn{2}{c}{(\kms)}
& \multicolumn{2}{c}{(\kms)}
& \multicolumn{2}{c}{(mK \kms)}
& \multicolumn{2}{c}{(\msun)}
&
\\
\hline
F2              & 22.3 & 1.7 & 346.7 & 1.5 & 18.0 & 1.5 &  1007.7 & 114.1 & 9.5 & $10^5$& $\llap{>}4.8$ \\
F3              & 17.0 & 1.0 & 332.1 & 0.6 &  8.8 & 0.6 &   374.2 &  31.9 & 3.5 & $10^5$ & $0.4$ \\
F4              &  8.7 & 0.8 & 322.7 & 1.0 &  9.7 & 1.0 &   210.7 &  28.8 & 2.0 & $10^5$& $0.2$ \\
F5              & 39.8 & 3.0 & 340.1 & 0.8 &  9.3 & 0.8 &   932.3 & 108.8 & 8.8 & $10^5$&  $\llap{>}4.4$ \\
F6              & 15.0 & 1.1 & 335.4 & 2.6 & 30.8 & 2.6 &  1154.7 & 131.0 & 10.9 & $10^5$ & $\llap{>}5.4$ \\
F6-1    & 19.0 & 1.2 & 316.8 & 1.1 & 11.0 & 1.0 &   523.6 &  58.1 & 4.9 & $10^5$ \\
F6-2  & 12.3 & 0.9 & 359.6 & 2.3 & 18.7 & 2.4 &   577.6 &  84.0 & 5.5 & $10^5$\\
\hline
\end{tabular}

\caption{Parameters of the \co\ profiles derived using Gaussian fits, derived molecular masses, and ratios of molecular to atomic mass for the five positions. For position F6 the results of a two-Gaussian fit are also given and are labelled F6-1 and F6-2. The molecular masses have been estimated using  CO(2-1)/CO(1-0) = 0.55 \citep{Charmandaris00} and a conversion factor  of  $\alpha_{\rm CO} = 1.5$ \msun\ (K km s$^{-1}$ pc$^2$)$^{-1}$.  \HI\ masses are derived from the data presented in \citet{Oosterloo05} using the same aperture as the APEX beam.}
 \label{tab:masses}
\end{table*}
\section{APEX observations}
Observations with the Atacama Pathfinder EXperiment (APEX) 12 m antenna were conducted between 7 and 10 April 2016, using the APEX-1 instrument with the XFFTS backend tuned to  230.5486 GHz, the frequency of  \co\  corresponding to the velocity of the gas in the outer filament ($V_{\rm hel}$$\sim$ 400 \kms). Five pointing positions were selected to cover different regions across the outer filament in order to sample very different conditions of the gas and related phenomena.  Their locations are shown in Fig.\ \ref{fig:pointing} and are listed in Table \ref{tab:obs}. Positions F3 and F4 were chosen to lie in the regularly rotating part of the \HI\ cloud in order to sample the quiescent \HI\ gas. To sample a location in the main ionised filament where the gas is clearly affected by the jet--cloud interaction,  we have chosen a position  (F6) where the ionised gas consists of at least two distinct kinematically and morphologically distinct components \citep{Santoro15b}. Finally, locations F2 and F5 lie just outside the \HI\ cloud at locations where some recent star formation has occurred.

The observations were done in good weather conditions (for these frequencies). The precipitable water vapour (PWV) was between 2.4 and 3.3 mm. The observations were made using 32768 channels covering a total band of 2.5 GHz ($\sim$2500 \kms) with a velocity resolution of 0.076 \kms. However, the final spectra were smoothed to 10 \kms\ bins. The data were reduced with the CLASS software from the Gildas package\footnote{http://www.iram.fr/IRAMFR/GILDAS} using the standard scripts provided by ESO.  From the individual scans of each of the XFFTS units a linear baseline was subtracted before adding all spectra.  The noise levels of the spectra are listed in Table \ref{tab:obs}. Table \ref{tab:masses} presents the parameters of the \co\ profiles for the five positions derived from Gaussian fits. At the frequency of our observations, the spatial  resolution of APEX  is  $\sim$$30^{\prime\prime}$ ($\sim$0.5 kpc)  and the gain is  39 Jy K$^{-1}$. 

\section{CO(2-1) in the region of the outer filament}

The first important result (Fig.\ \ref{fig:profiles} and Table  \ref{tab:masses}) is that we  detect \co\ at {\em all  five locations}  spread over the {\em entire region}: in the \HI\ cloud (F3 and F4), in the main region of ionised gas (F6), but also at locations in between the \HI\ cloud and the ionised gas (F2 and F5). The first direct consequence  is that the molecular gas is the only gas tracer that is present inside the \HI\ cloud, in the ionised filament, and in between the two, while  \HI\ and ionised gas  are detected  in mutually excluding locations. This underlines, as  in other cases of jet--cloud interactions, that the molecular gas is a crucial component in such phenomena.  

Although limited by the small number of pointings,  a trend can be seen in the width of the \co\ profile.  The widths increase going from  west to east, from inside the \HI\ cloud to outside it, and finally to the region of ionised gas.  This mirrors what is  seen in the \HI\  (which has narrow profiles only $\sim $20\kms\ wide;  \citealt{Oosterloo05}) and  the ionised gas (with much more complex kinematics with line widths $> 100$ \kms; \citealt{Morganti91,Santoro15b}). The velocities and widths of the profiles F3, F4, and F5 are consistent with those of the regularly rotating part of the \HI\ cloud and with what was obtained by \citet{Charmandaris00} for a region near F5. In contrast,  the detections towards and in the region of the ionised filament (F2 and F6) are much broader and show gas at velocities corresponding to the high-velocity, disturbed components  of the ionised gas (see \citealt{Santoro15a}) and are deviating from the extrapolation of the regularly rotating \HI. 
The profile at F6 is broadest  with a FWHM = 72 \kms,  is asymmetric and  shows some indications of having multiple components.   The possible presence of multiple components at position F6 may be related  to the presence of different kinematical components seen in the ionised gas, in which case it would imply a close connection between the two phases.

It has been noted by several groups  that the star formation rate in the outer filament is low  and that the rate at which the available cold gas is turned into stars is also low \citep{Mould00,Rejkuba04,Oosterloo05,Salome16}. \citet{Salome16}  suggest that  turbulence due to kinetic energy injection from the AGN jet leads to molecular gas reservoirs not forming stars efficiently and to quenching of star formation. 
The differences in line widths we detect  do indeed suggest that turbulence is increased due to the jet--cloud interaction. 
However, star formation only occurs in the region {\em outside} the \HI\ cloud, albeit at low rates,  while  it seems to be completely absent in the region {\em inside} the \HI\ cloud.  It therefore seems that the star formation that is occurring is not being quenched, but instead is stimulated by the passage of the radio jet. It is, on the other hand, rather puzzling that no star formation is occurring in the \HI\ cloud where there are no indications that conditions are particularly adverse to star formation. The relative locations of the ionised gas and the \HI\ are also suggestive that the jet plays a role in the star formation (see Fig.\ 1). The general possibility of jet-induced star formation has been considered by theoretical models \citep{Mellema02,Fragile04,Gaibler12}, but  no  study has been made for conditions similar to those of the outer filament. 

In order to derive  molecular masses from our \co\ detections, we have made assumptions about the CO(2-1)/CO(1-0) ratio and the conversion factor CO-to-H$_2$.
For the former we have assumed the factor obtained by \citet{Charmandaris00}  for their Shell S1 region, which is located close to F5: CO(2-1)/CO(1-0) $ = 0.55$. This may not be correct for  all  regions covered by our observations and the differences in total line flux between the positions inside and outside the \HI\ cloud could  partly be due to  different ratios at different locations with different conditions. However, until observations of more transitions are available, it is difficult to apply a varying  ratio.

The choice of the conversion factor for  CO to H$_2$ requires some considerations.
\citet{Charmandaris00} have used a standard value, i.e.\ $\alpha_{\mathrm{CO}} = 4.6$ \msun\ (K km s$^{-1}$ pc$^2$)$^{-1}$.  However, more conservative assumptions have been used for detections in  radio AGN.  For example, \citet{Evans05} and \citet{Smolcic11} have used  $\alpha_{\mathrm{CO}} = 1.5$ \msun\ (K km s$^{-1}$ pc$^2$)$^{-1}$ for CO detected in radio galaxies. 
In Table \ref{tab:masses} we list the estimated molecular gas masses  using the more conservative assumption. Nevertheless, the relatively low metallicity of the gas may suggest that larger conversion factors would be more realistic, resulting in higher values for the H$_2$ masses. However, an even more complex scenario is conceivable where differences in  conversion factors exist between different regions owing to their very different physical conditions (e.g.\ velocity structure of the gas, influence of the jet).  Ignoring such complications (which can only be addressed by more detailed observations) and using the more conservative assumption, the masses of the molecular gas range between $0.2 \times 10^6$ and $1.1 \times 10^6$ \msun\ and, like  the kinematics, they show differences  between the regions coincident with the \HI, having the lower H$_2$ masses, and the ones outside, showing higher H$_2$ masses. 
Again we want to note that the apparent difference in masses between inside and outside the \HI\ cloud could be   due in part to not taking different physical conditions into account. A direct comparison of  the masses  derived above with those of the detections of \citet{Charmandaris00} is not possible because of the larger area covered by their mosaic observation. However, the peak fluxes and widths  of the profiles shown by \citet{Charmandaris00} are very similar to those of our spectra. 

In Table  \ref{tab:masses} we also present  the ratio between molecular and atomic gas masses, $M_{\rm H_2}/M_{\rm \HI}$, where we have derived the \HI\ masses from the data of \citet{Oosterloo05} using  an aperture of the same size as the APEX beam for each position. We find that for the two regions coincident with the \HI\ cloud this ratio is $\sim$0.3 while we find a much higher lower limit ($>$4) for the other regions. 
\citet{Charmandaris00}  derived a value $M_{\rm H_2}/M_{\rm \HI} = 0.8$ for their detection in the \HI\ cloud. Considering the difference in the CO-to-H2 conversion factor used, our results appear to be consistent with theirs. Despite the uncertainties in the conversion factors used, the difference in $M_{\rm H_2}/M_{\rm \HI}$ we find is too large to be  due to  these uncertainties alone. In fact, the distinctive factor for the difference in mass ratios is the lack of \HI\  at some locations, more so than the differences in molecular mass. We therefore conclude that the large contrast in mass ratios suggests that the atomic hydrogen is much more affected by the jet--cloud interaction than the molecular gas.

\section{Origin and fate of the molecular gas}

One of the main puzzles of our results is the overall presence of molecular gas, despite the large range of  conditions characterising this region near Cen A. 
One  scenario to explain the observations is that   molecular  clouds originally residing inside the \HI\ cloud are not much affected by the jet--cloud interaction, while  the more tenuous atomic hydrogen of the \HI\ cloud is. The molecular gas in the \HI\ cloud is likely to be in denser clumps which   would, in this scenario, \emph{manage to survive} the effects of the passing jet. On the other hand, the \HI\ itself, being less dense, {\em is} affected by the jet and is blown away more easily. The \HI\ is ionised while flowing downstream, resulting in  high values of  $M_{\rm H_2}/M_{\rm \HI}$. The overall kinematics of the molecular clouds would be very similar to that of the \HI\ cloud (or the extrapolation of it to the regions devoid of \HI). The increased profile widths of the molecular gas down the flow would mean that, although this gas survives, increased turbulence is somehow induced by the interaction. It has been suggested that the increased turbulence causes the low star formation efficiency in the outer filament \citep{Salome16}.
This scenario would explain the relative distribution of the different  gas phases, the very different ratios $M_{\rm H_2}/M_{\rm \HI}$, and some of the kinematics. This scenario may occur if, as suggested by the overall kinematics of the ionised gas, only a mild interaction occurs between a slowly moving radio jet  and the ISM.

An alternative scenario is similar to what is proposed for other cases of jet--cloud interactions (e.g.\ \citealt{Tadhunter14}). Here it is assumed that \emph{all the gas} (atomic and molecular) gets photo- or shock-ionised when entering the region of the radio jet.  However,  denser clumps rapidly reform   in the wake of the interaction and cool very efficiently, forming new clouds of molecular gas downstream which ultimately start forming stars.  In this scenario  \HI\  is only  a very short transient phase in the cooling of the gas from ionised to molecular and its column densities are below the detection limit. As a result, the molecular gas is the most massive component in this region because it is the end product of a fast cooling process and it accumulates over time, giving high values for $M_{\rm H_2}/M_{\rm \HI}$. Since the molecular clumps form from the ionised gas, the kinematics of the molecular gas would mirror that of the ionised gas and less that of the atomic gas.

With the available data, it is not easy to distinguish between the two scenarios.  
The large profile widths measured outside the \HI\ cloud, as well as the possible presence of multiple components in the  same velocity range as those seen in the ionised gas,  could suggest that  the kinematics of the  H$_2$ at these locations is similar to that of the ionised gas, which  covers a larger velocity range than that of the \HI\ and which consists of  separate components. This may indicate that the molecular gas  has the momentum of the interaction  encoded in it, in a similar way to the ionised gas, which would argue for the second scenario. On the other hand, the narrow width of profile F5 just outside the \HI\ cloud matches that of F3 and F4 inside it and would better fit the first scenario of exposed, pre-existing cold H$_2$ cores. It might well be that one scenario is more important in one region and the other scenario at other locations, depending on the local strength and geometry of the jet--cloud interaction. 

More extensive observations of the molecular gas, including higher resolution imaging, will be required to study this further. Imaging the region at high spatial resolution with ALMA  would allow us to investigate the turbulence and velocity structure of the gas on the scales of  individual clouds  and as a function of position, at very similar resolution to the optical data. This would make it possible to obtain a detailed view of the connection between the different gas phases and of the processes which occur in the jet--cloud interaction in the outer filament of Cen A.

\noteaddname\
During the evaluation of this manuscript, we became aware of a manuscript
by Salome et al. (arXiv:1605.05986) that also discusses CO(2-1) APEX 
observations of this region of Centaurus A. It mostly addresses distinct
issues but is consistent/compatible with our conclusions.

\begin{acknowledgements}
We would like to thank Carlos de Breuck and Theresa Nilsson for their support for the observations. This publication is based on data acquired with the Atacama Pathfinder EXperiment (APEX) which is a collaboration between the Max-Planck-Institut f\"{u}r Radioastronomie,  ESO, and the Onsala Space Observatory. We acknowledge the use of the GILDAS software for the  data reduction. The research leading to these results has received funding from the European Research Council under the European Union's Seventh Framework Programme (FP/2007-2013)/ERC Advanced Grant RADIOLIFE-320745.
\end{acknowledgements}

%
%

\end{document}